\documentclass[prd,twocolumn,floatfix,preprintnumbers,showpacs]{revtex4}
\usepackage{graphicx}
\usepackage{dcolumn}
\usepackage{bm}

\def\lsim{\mathrel{\mathop
  {\hbox{\lower0.5ex\hbox{$\sim$}\kern-0.8em\lower-0.7ex\hbox{$<$}}}}}
\def\gsim{\mathrel{\mathop
  {\hbox{\lower0.5ex\hbox{$\sim$}\kern-0.8em\lower-0.7ex\hbox{$>$}}}}}

\begin{document}

\newcommand{\half}{{1\over2}}
\newcommand{\nad}{n_{\rm ad}}
\newcommand{\niso}{n_{\rm iso}}
\newcommand{\fiso}{f_{\rm iso}}
\newcommand{\ii}{\'{\'i}}
\newcommand{\Ocdm}{\Omega_{\rm cdm}}
\newcommand{\ocdm}{\omega_{\rm cdm}}
\newcommand{\OM}{\Omega_{\rm M}}
\newcommand{\OB}{\Omega_{\rm B}}
\newcommand{\oB}{\omega_{\rm B}}
\newcommand{\OL}{\Omega_\Lambda}
\newcommand{\cltt}{C_l^{\rm TT}}
\newcommand{\clte}{C_l^{\rm TE}}
\newcommand{\hatR}{{\cal R}}
\newcommand{\hatS}{{\cal S}}

\input epsf

\title{Bounds on isocurvature perturbations from CMB and LSS data}
\author{Patrick Crotty,$^1$ Juan Garc{\'\i}a-Bellido,$^2$ Julien 
Lesgourgues,$^{1,3}$ and Alain Riazuelo$^4$}
\affiliation{$^1$Laboratoire de Physique Th\'eorique LAPTH, F-74941
Annecy-le-Vieux Cedex, France\\
$^2$Departamento de F\'\i sica Te\'orica \ C-XI, Universidad
Aut\'onoma de Madrid, Cantoblanco, 28049 Madrid, Spain\\
$^3$TH-Division CERN, CH-1211 G\`eneve 23, Switzerland\\
$^4$Service de Physique Th\'eorique CNRS, CEA/Saclay F-91191, 
Gif-sur-Yvette Cedex, France}
\pacs{98.80.Cq}
\begin{abstract}
We obtain very stringent bounds on the possible cold dark matter,
baryon and neutrino isocurvature contributions to the primordial
fluctuations in the Universe, using recent cosmic microwave background
and large scale structure data. In particular, we include the measured
temperature and polarization power spectra from WMAP and ACBAR, as
well as the matter power spectrum from the 2dF galaxy redshift survey.
Neglecting the possible effects of spatial curvature, tensor
perturbations and reionization, we perform a Bayesian likelihood
analysis with nine free parameters, and find that the amplitude of the
isocurvature component cannot be larger than about 31\% for the cold
dark matter mode, 91\% for the baryon mode, 
76\% for the neutrino density mode, and 60\% for the
neutrino velocity mode, at 2-$\sigma$, for uncorrelated models. On the
other hand, for correlated adiabatic and isocurvature components, the
fraction could be slightly larger. However, the cross-correlation
coefficient is strongly constrained, and maximally
correlated/anticorrelated models are disfavored. This puts strong
bounds on the curvaton model, independently of the bounds on
non-Gaussianity.
\end{abstract}

\maketitle

{\bf Introduction}. Thanks to the tremendous developments in
observational cosmology during the last few years, it is possible to
speak today of a Standard Model of Cosmology, whose parameters are
known within systematic errors of just a few percent. Moreover, the
recent measurements of both temperature and polarization anisotropies
in the cosmic microwave background (CMB) has opened the possibility to
test not only the basic paradigm for the origin of structure, namely
inflation, but also the precise nature of the primordial fluctuations
that gave rise to the CMB anisotropies and the density perturbations
responsible for the large scale structure (LSS) of the Universe.

The simplest realizations of the inflationary paradigm predict an
approximately scale invariant spectrum of adiabatic and Gaussian
curvature fluctuations, whose amplitude remains constant outside the
horizon, and therefore allows cosmologists to probe the physics of
inflation through observations of the CMB anisotropies and the LSS
matter distribution. However, this is by no means the only
possibility.  Multiple-field inflation predicts that, together with
the adiabatic component, there should also be an entropy or
isocurvature
perturbation~\cite{Linde:1985yf,Polarski:1994rz,Gordon:2000hv},
associated with fluctuations in number density between different
components of the plasma before decoupling, with a possible
statistical correlation between the adiabatic and isocurvature
modes~\cite{Langlois:1999dw}.  Baryon and cold dark matter (CDM)
isocurvature perturbations were proposed long
ago~\cite{Efstathiou:1986} as an alternative to adiabatic
perturbations.  Recently, two other modes, neutrino isocurvature
density and velocity perturbations, have been added to the
list~\cite{Rebhan:1994zw}. Moreover, it is well known that
entropy perturbations seed curvature perturbations outside the
horizon~\cite{Polarski:1994rz,Gordon:2000hv}, so it is possible that a
significant component of the observed adiabatic mode could be strongly
correlated with an isocurvature mode. Such models are generically
called {\em curvaton models}~\cite{Lyth:2001nq,Lyth:2002my}, and are
now widely studied as an alternative to the standard
paradigm. Furthermore, isocurvature modes typically induce
non-Gaussian signatures in the spectrum of primordial perturbations.

In this Letter we present very stringent bounds on the various
isocurvature components, coming from the temperature power spectrum
and temperature-polarization cross-correlation recently measured by
the WMAP satellite~\cite{WMAP}; from the small-scale temperature
anisotropy probed by ACBAR~\cite{ACBAR}; and from the matter power
spectrum measured by the 2-degree-Field Galaxy Redshift Survey
(2dFGRS)~\cite{2dFGRS}. We do not use the data from Lyman-$\alpha$
forests, since they are based on non-linear simulations carried under
the assumption of adiabaticity.  We will not assume any specific model
of inflation, or any particular mechanism to generate the
perturbations (late decays, phase transitions, cosmic defects, etc.),
and thus will allow all five modes -- adiabatic (AD), baryon
isocurvature (BI), CDM isocurvature (CDI), neutrino isocurvature
density (NID) and neutrino isocurvature velocity (NIV) -- to be
correlated (or not) among each other, and to have arbitrary
tilts. However, we will only consider the mixing of the adiabatic mode
and one of the isocurvature modes at a time. This choice has the
advantage of restricting the number of free parameters, and takes
into account the fact that most of the proposed mechanisms for the
generation of isocurvature perturbations lead to only one mode. The
first bounds on isocurvature perturbations assumed uncorrelated
modes~\cite{Stompor:1995py}, but recently also correlated ones were
considered~\cite{Trotta:2001yw,Amendola:2001ni,Gordon:2002gv,Valiviita:2003ty}.

The present analysis neglects the possible effects of spatial
curvature, tensor perturbations and reionization. Therefore, each
model is described by nine parameters: the cosmological constant
$\OL$, the baryon density $\oB=\OB h^2$, the cold dark matter density
$\ocdm=\Ocdm h^2$, the overall normalisation $A$, the isocurvature
mode relative amplitude $\alpha$ and correlation $\beta$, the
adiabatic and isocurvature tilts ($\nad$, $\niso$), and finally a free
bias $b$ associated to the 2dF power spectrum. We generate a
5-dimensional grid of models ($A$, $\alpha$, $\beta$, and $b$ are not
discretized) and perform of Bayesian analysis in the full
9-dimensional parameter space. At each grid point, we store some $C_l$
values in the range $0 < l < 1800$ and some $P(k)$ values in the range
probed by the 2dF data. The likelihood of each model is then computed
using the software or the detailed information provided on the
experimental websites, using 1398 points from WMAP, 11 points from
ACBAR and 32 points from the 2dFGRS. For parameter values which do not
coincide with grid points, our code first performs a cubic
interpolation of each power spectrum, accurate to better than one
percent, and then computes the likelihood of the corresponding model.
From the limitation of our grid, we impose the flat prior $\niso >
0.6$ (as preferred by inflation). The other grid ranges are wide
enough in order not to affect our results.

For the theoretical analysis, we will use the notation and some of the
approximations of Ref.~\cite{Amendola:2001ni}. For instance, the
power spectra of adiabatic and isocurvature perturbations, as well as
their cross-correlation, are parametrized with 
three amplitudes and two spectral indices,
\begin{eqnarray}\nonumber
\Delta_{\cal R}^2(k)\!&\equiv&\!
{k^3\over2\pi^2}\langle\hatR^2\rangle = {\cal A}^2\,\Big({k\over k_0}\Big)^{\nad-1}\,,\\[2mm]
\Delta_{\cal S}^2(k)\!&\equiv&\!
{k^3\over2\pi^2}\langle\hatS^2\rangle = {\cal B}^2\,\Big({k\over k_0}\Big)^{\niso-1}\,,\\[2mm]
\Delta_{{\cal R}{\cal S}}^2(k)\!&\equiv&\!{k^3\over2\pi^2}\nonumber
\langle\hatR\hatS\rangle = {\cal A}\,{\cal B}\,
\cos\Delta\ \Big({k\over k_0}\Big)^{(\nad+\niso)/2-1}\,.
\end{eqnarray}
where $k_0$ is an arbitrary pivot scale.
We also assume that the correlation coefficient $\cos\Delta$ is
scale-independent (Note that in Ref.~\cite{Valiviita:2003ty} this
assumption was relaxed). In order to evaluate the temperature and
polarization anisotropies, one has to calculate the radiation transfer
functions for adiabatic and isocurvature perturbations and compute the
total angular power spectrum as~\cite{Amendola:2001ni}
\begin{equation}
C_l = C_l^{\rm ad} + B^2\,C_l^{\rm iso} + 
2B\,\cos\Delta\,C_l^{\rm corr}\,,
\end{equation}
where 
$B$ is the entropy to curvature perturbation ratio during radiation
domination, $B={\cal S}/{\cal R}$. We
will use here a slightly different notation, used before by other
groups~\cite{Langlois:1999dw,Stompor:1995py}, where 
\begin{equation}\label{alphanotation}
C_l = (1-\alpha)\,C_l^{\rm ad} + \alpha\,C_l^{\rm iso} + 2\beta\,
\sqrt{\alpha(1-\alpha)}\,C_l^{\rm corr}\,.
\end{equation}
The two notations are related by
\begin{equation}
\alpha = B^2/(1+B^2)\,, \hspace{1cm} \beta = \cos\Delta\,.
\end{equation}
This notation has the advantage that the full parameter space of
$(\alpha,\ 2\beta\sqrt{\alpha(1-\alpha)})$ is contained within a
circle of radius $1/2$. The North and South rims correspond to fully
correlated ($\beta=+1$) and fully anticorrelated ($\beta=-1$)
perturbations, with the equator corresponding to uncorrelated
perturbations ($\beta=0$).  The East and West correspond to purely
isocurvature and purely adiabatic perturbations, respectively. Any
other point within the circle is a particular admixture of adiabatic
and isocurvature modes.

\begin{figure*}[ht]
\includegraphics[angle=-90,width=7cm]{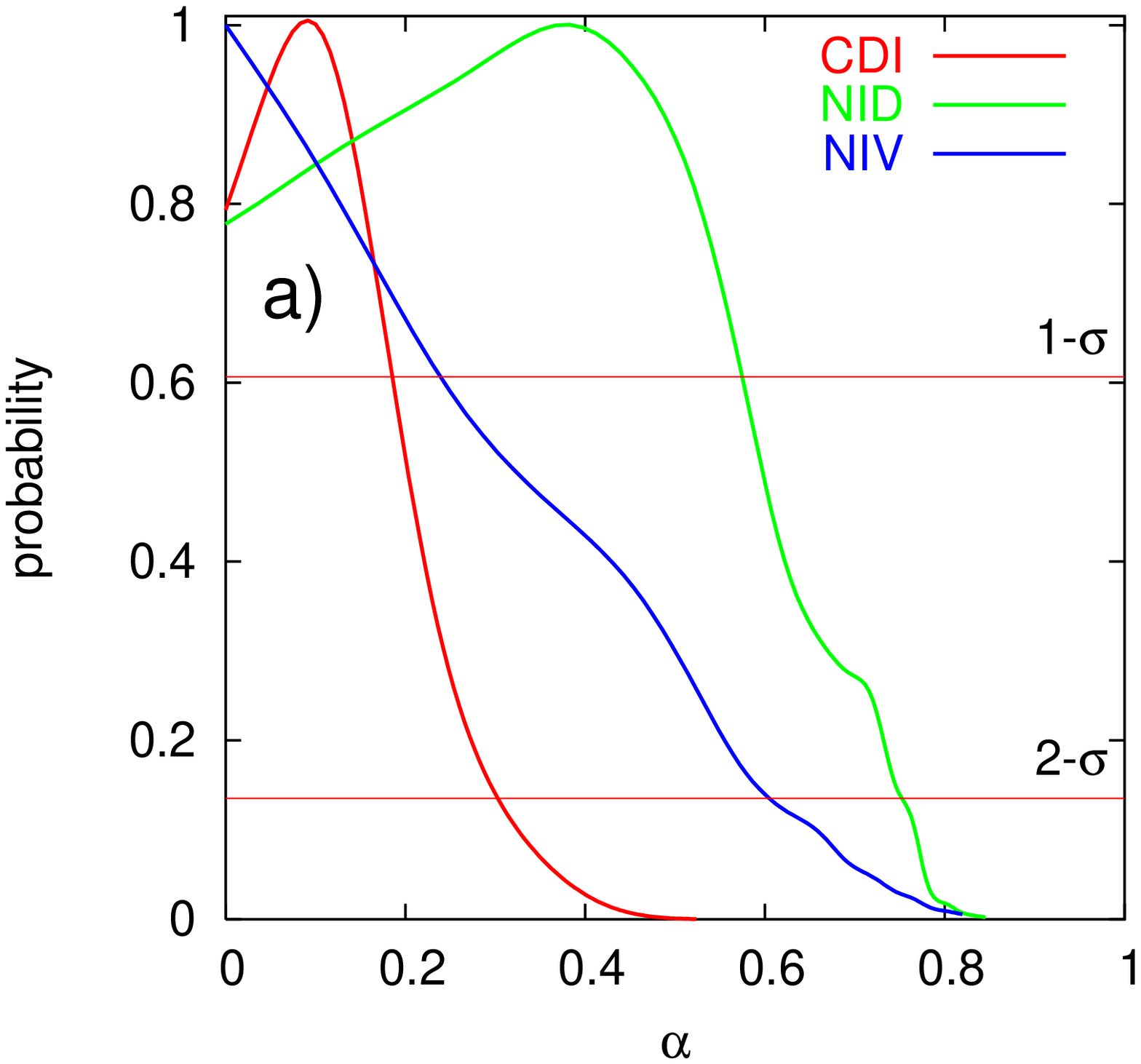}
\includegraphics[angle=-90,width=7cm]{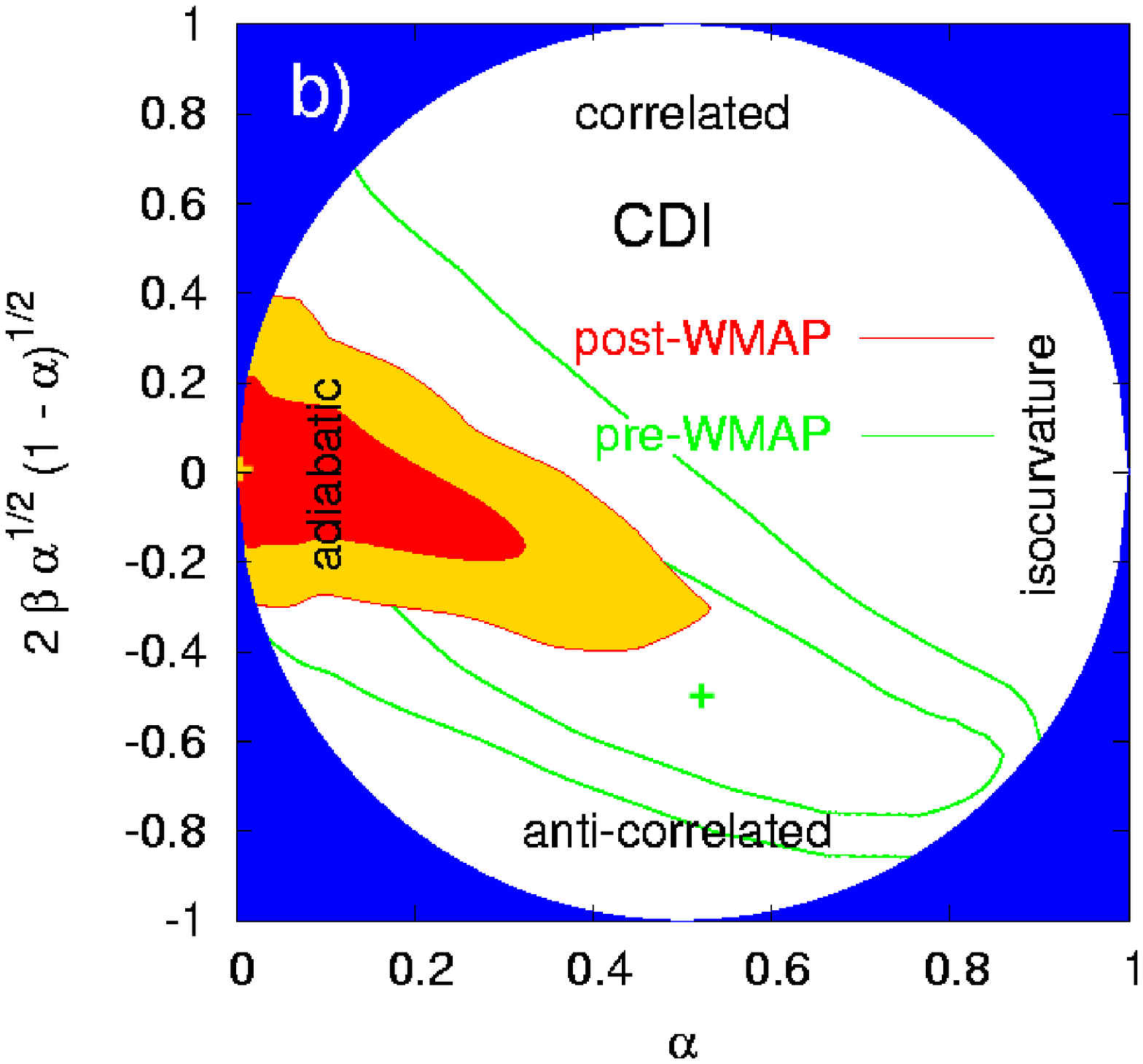}
\includegraphics[angle=-90,width=7cm]{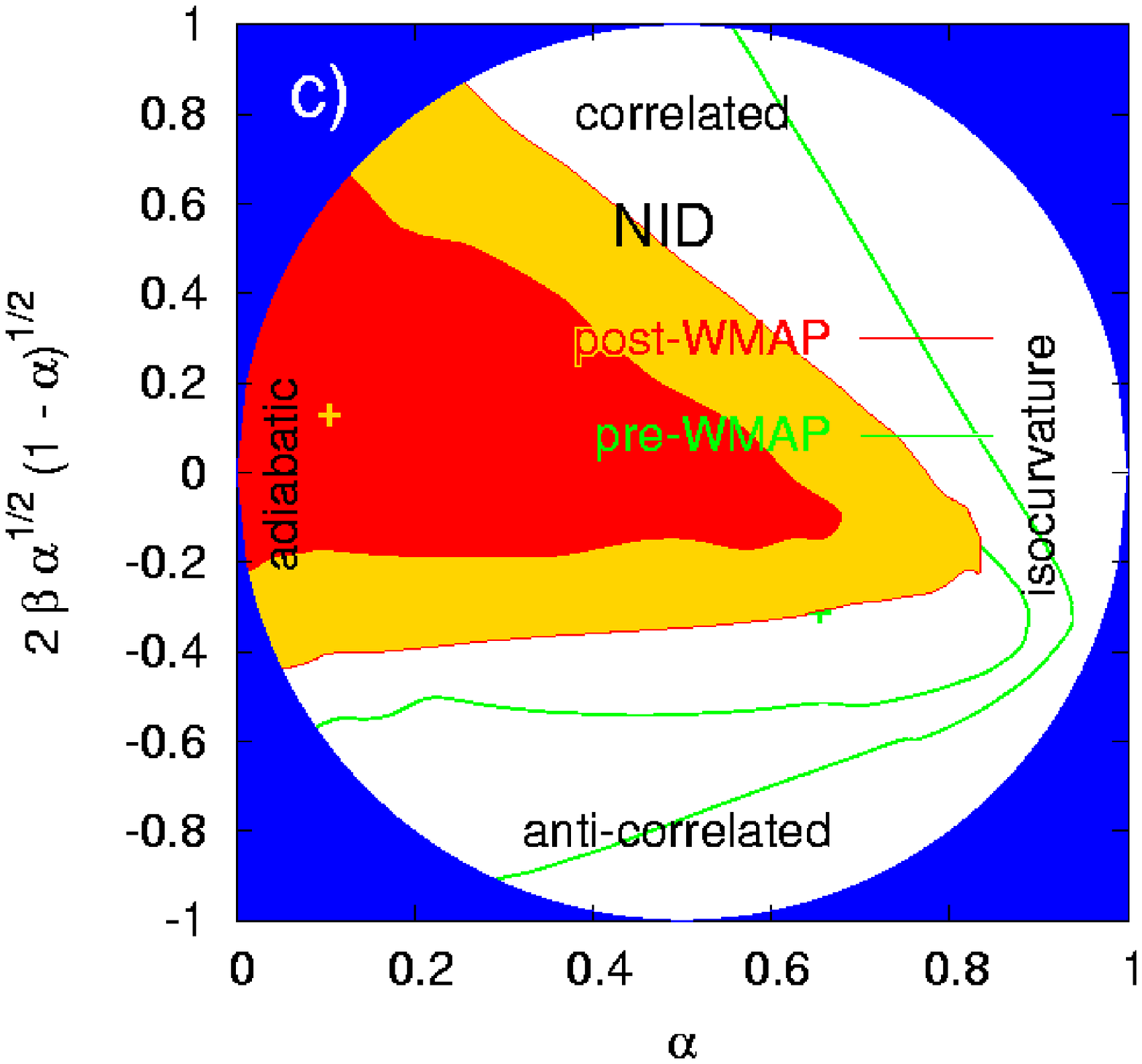}
\includegraphics[angle=-90,width=7cm]{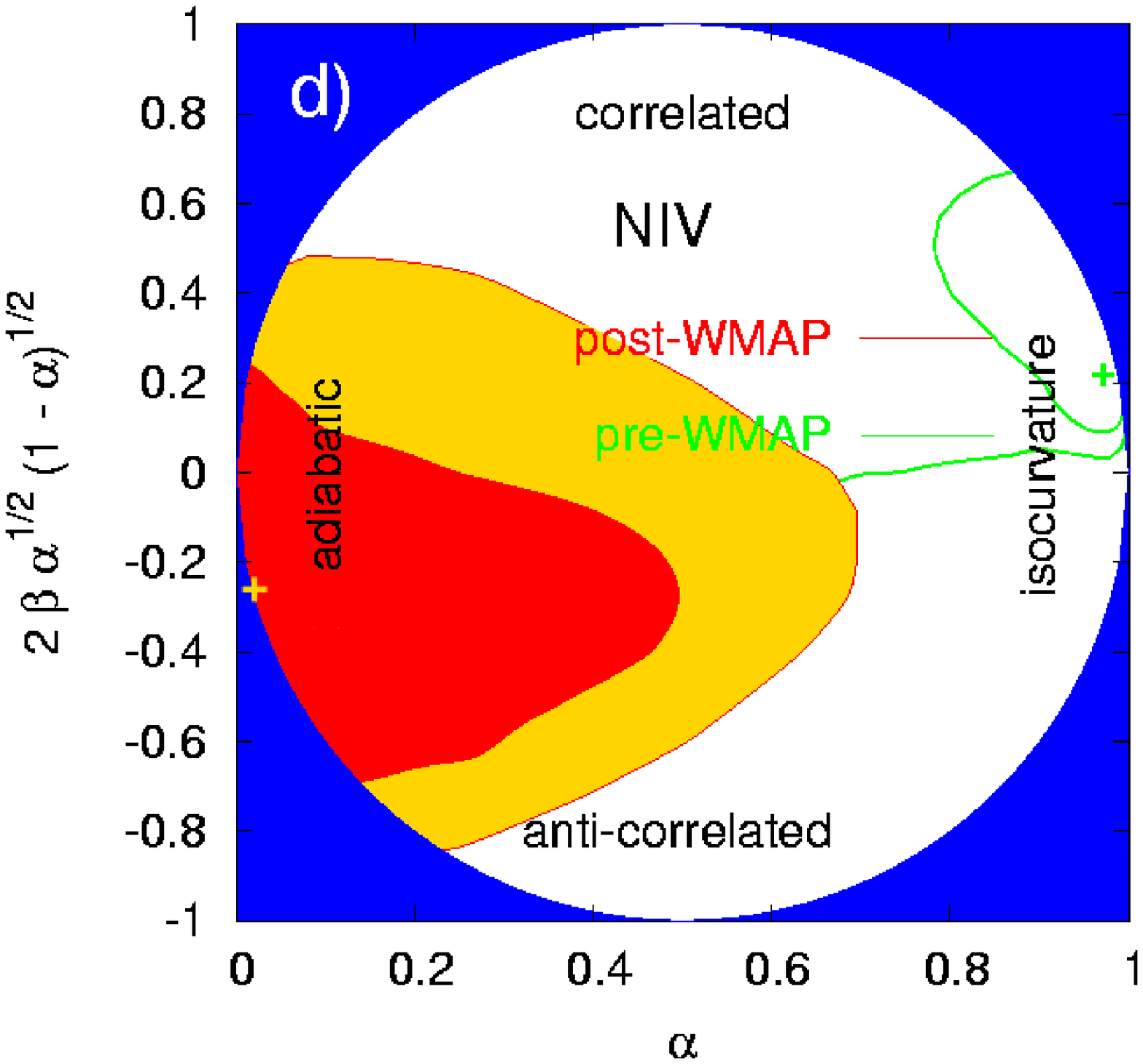}
\caption{\label{alpha} a) the likelihood function of the isocurvature
fraction $\alpha$, for three different types of uncorrelated
isocurvature modes (i.e., with the prior $\beta = 0$); b) the 1 and
2-$\sigma$ contours of $\alpha$ and the cross-correlated mode
coefficient $2 \beta\sqrt{\alpha(1-\alpha)}$, for the CDM
isocurvature mode: the small (red) contours are based on all the data,
with one flat prior $\niso > 0.6$,
while the large (green) ones show the situation before WMAP,
with an additional prior $\omega_B < 0.037$; c) same
as b) for NID; d) same as b) for NIV.
The marginalization is approximated by a maximum likelihood
fit of the other parameters for each pair of values $(\alpha, \beta)$.
}
\end{figure*}

In order to compute the $C_l$
coefficients of the temperature and polarization power spectra, as
well as the matter spectra $P(k)$, 
we have used a CMB code developed by one of us (A.R.) which coincides,
within 2\% errors, with the values provided by CMBFAST for AD and CDI
modes, and also includes the neutrino isocurvature modes as well
as the cross-correlated power spectra. Note that the code defines
the tilt of each power spectrum with respect to a pivot
scale $k_0$ corresponding the present value of the Hubble radius, while
CMBFAST uses $k_0 = 0.05$~Mpc$^{-1}$: so, the comparison of our
results with those of Refs.~\cite{WMAP} for the CDI mixed model 
is not straightforward.

\begin{table}[h]
\caption{The one-dimensional 2-$\sigma$ ranges on 
the isocurvature mode coefficients for the various models,
uncorrelated (middle column) and correlated (right column).}
\begin{tabular}{|c| c|cr|}
\hline
model & $\alpha$ &
$\alpha$ &  $2 \beta\, [\alpha (1 - \alpha)]^{1/2}$\cr 
\hline \hline
CDI & $< 0.31$ & $< 0.47$ &\hspace{2mm} 
$-0.31$ \ to \ $0.31$ \cr
BI & $< 0.91$ & $< 0.95$ &\hspace{2mm} 
$-0.80$ \ to \ $1.00$ \cr
NID & $< 0.76$ & 
$< 0.77$ &\hspace{2mm}  
$-0.30$ \ to \ $0.78$ \cr
NIV & $< 0.60$ & $< 0.60$ & \hspace{2mm} 
$-0.77$ \ to \ $0.35$ \cr
\hline
\end{tabular}
\label{table2}
\end{table}

{\bf Results}. We now describe the different bounds, summarized 
in Table~I. For each specific mode,
the probability distribution for $\alpha$ (uncorrelated case)
and the likelihood contours in the $(\alpha,\,\beta)$ 
plane (correlated case) are shown on Fig.~1.
The best-fitting parameter set for each model is given in Table~II.

{\it CDM isocurvature}. Assuming no correlation between the adiabatic
and CDI modes, we find an upper bound $\alpha < 0.31$ at 95\% c.l..
The deviation from a pure adiabatic model does {\it not} improve the
goodness-of-fit, since the minimum $\chi^2$ value goes from 1478.8 to
1478.3, while the number of degrees of freedom decreases from 1435 to
1433. The situation does not change significantly when we include a
possible correlation: the minimum $\chi^2$ is still around 1478.0, and
the individual 2-$\sigma$ bounds on the coefficients read $\alpha <
0.47$ and $-0.31 < 2\beta \sqrt{\alpha(1 - \alpha)} < 0.31$ at 95\%
c.l..
Most of the constraints on this model come from the
WMAP TT power spectrum. For comparison, we repeated our analysis
replacing WMAP and ACBAR by the Wang et al.~\cite{Wang:2002rt} CMB
data compilation, which includes all the pre-WMAP temperature
data. In that case, the results exhibit a wide parameter degeneracy
(as indicated by previous studies
\cite{Trotta:2001yw,Amendola:2001ni}), and a large fraction of
anti-correlated isocurvature modes cannot be excluded, up to
$(\alpha,\,\beta) = (0.9,-1)$, provided one allows for a very large
baryon fraction.  Therefore, we conclude that the WMAP TT data is very
powerful in constraining the isocurvature fraction. On the other hand,
we checked explicitly that the effect of the WMAP TE power spectrum is
not very strong with respect to that of TT.

In the curvaton scenario, the case in which the CDM is created before
the curvaton decay while the curvature perturbations are small,
corresponding to $(\alpha,\,\beta)= (0.9,\,-1)$ in our
notation~(\ref{alphanotation}), is completely excluded, as already
emphasized in Ref.~\cite{Gordon:2002gv}. The case in which the CDM is
created by the decay of the curvaton leads to $\beta=1$; with such a
prior, we obtain a sharp 2-$\sigma$ bound: $\alpha < 0.04$ or, in the
notation of Ref.~\cite{Amendola:2001ni}, $B <0.2$ (the authors of
Ref.~\cite{Gordon:2002gv} obtain a more restrictive bound $B < 0.43\
\OB/\OM \sim 0.1$, presumably because they use the curvaton--motivated
assumption $\niso = \nad$).  We can rewrite this bound as a constrain
on the fraction $r=\Omega_{\sigma,\,{\rm decay}}$ since, in these models,
$B=3(1-r)/r$.  In our case, $B <0.2$ implies $1-r<0.0625$.

\begin{table}[ht]
\caption{The best fit values of the parameters for the adiabatic (AD),
uncorrelated (CDI, etc.), and correlated (c-CDI, etc.) isocurvature 
models, with the corresponding $\chi^2$ and number of degrees of 
freedom $\nu$. For the NIV uncorrelated model, the best-fit occurs 
for $\alpha=0$ (i.e., purely adiabatic).}
\begin{tabular}{|c|ccccccc|c|}
\hline
model & $\oB$ & $\ocdm$ & $\OL$ 
& $\nad$ & $\niso$ & $\alpha$ & $\beta$ 
& $\chi^2/\nu$ \cr \hline \hline
AD  
& $0.021$ & $0.12$ & $0.70$ & $0.95$ & $-$ & $-$ & $-$ 
& $1478.8/1435$ \cr
\hline
CDI
& $0.023$ & $0.12$ & $0.73$ & $0.99$ & $1.02$ & $0.10$ & $-$ 
& $1478.3/1433$ \cr
BI
& $0.023$ & $0.12$ & $0.73$ & $0.99$ & $1.02$ & $0.72$ & $-$ 
& $1478.3/1433$ \cr
NID
& $0.023$ & $0.12$ & $0.73$ & $0.99$ & $0.95$ & $0.37$ & $-$ 
& $1478.2/1433$ \cr
NIV
& $0.021$ & $0.12$ & $0.70$ & $0.95$ & $-$ & $0$ & $-$ 
& $1478.8/1433$ \cr
\hline
c-CDI
& $0.022$ & $0.12$ & $0.71$ & $0.97$ & $1.23$ & $0.001$ & $1$ 
& $1478.0/1432$ \cr
c-BI
& $0.022$ & $0.12$ & $0.71$ & $0.97$ & $1.23$ & $0.03$ & $1$ 
& $1478.0/1432$ \cr
c-NID
& $0.022$ & $0.12$ & $0.73$ & $0.97$ & $1.04$ & $0.10$ & $0.26$ 
& $1477.7/1432$ \cr
c-NIV
& $0.021$ & $0.12$ & $0.71$ & $0.95$ & $0.71$ & $0.03$ & $-1$ 
& $1477.5/1432$ \cr
\hline
\end{tabular}
\label{table1}
\end{table}

{\it Baryon isocurvature}. The case of baryon isocurvature modes is
qualitatively similar to that of CDI modes, since the spectra
are simply rescaled by a factor $\Omega_{\rm B}^2/\Omega_{\rm cdm}^2$
($\Omega_{\rm B}/\Omega_{\rm cdm}$) for the isocurvature power
(cross-correlation) components: thus, significantly larger values of
$\alpha$ will be allowed in the baryon case.  Some approximate results for
the BI modes could be deduced from the CDI results by an overall
rescaling of the $\alpha$ parameter. However, we performed an exact
analysis, and found the following bounds: $\alpha < 0.91$
(uncorrelated case), or $\alpha < 0.95$ and $-0.8 < 2
\beta\sqrt{\alpha (1 - \alpha)} < 1$ (correlated case).  In each
case, the minimum $\chi^2$ is, by definition, the same as for the CDI
case. We provide the best-fit parameters in Table I, but for
brevity we do not show this case in the figures.

In the curvaton scenario, the case in which the baryon number is
created before the curvaton decay while the curvature perturbations
are small, corresponding to $(\alpha,\,\beta)= (0.9,\,-1)$, is also
excluded at several $\sigma$'s, as previously found by in
Ref.~\cite{Gordon:2002gv}.  The case in which the baryon number is
created by the decay of the curvaton predicts $\beta=1$; we then find
$\alpha < 0.5$, or $B < 1 $.

{\it Neutrino density isocurvature}. The 2-$\sigma$ bound for the
uncorrelated case is $\alpha < 0.76$, while for the correlated
adiabatic and neutrino mode we obtain $\alpha < 0.77$ and $-0.30 < 2
\beta\sqrt{\alpha (1 - \alpha)} < 0.78$.  
The best-fitting models for the two cases 
are still extremely close to the adiabatic model, leading to no significant
improvement in the likelihood, $\Delta\chi^2 = 0.6$ vs. $\Delta\chi^2
= 1.1$, respectively. Note that with the pre-WMAP CMB data plus the
2dFGRS power spectrum, significantly larger fractions, up to $(\alpha,\,\beta)
= (0.9,\,-0.6)$, of correlated NID modes were then allowed, but are
now excluded with great confidence.

The authors of Ref.~\cite{Lyth:2002my} discuss an interesting
mechanism under which the curvaton could generate some fully
correlated or anticorrelated NID modes, through inhomogeneities in the
neutrino/antineutrino asymmetry parameter. However, our bounds for
these cases are very restrictive, $\alpha < 0.18$ (correlated), or
$\alpha < 0.02$ (anticorrelated). In the notation of
Ref.~\cite{Amendola:2001ni}, this is equivalent to $B < 0.47$ or $B <
0.14$, respectively.

{\it Neutrino velocity isocurvature}. 
Again, the presence of such
a mode is not significantly favored: in the uncorrelated case, the
best-fitting model is still the pure adiabatic one ($\alpha=0$), while
in the correlated case the $\chi^2$ improvement is only $\Delta \chi^2
= 1.3$.  The impact of WMAP on these bounds is spectacular. Using only
the pre-WMAP CMB data compilation and the 2dFGRS power spectrum, we
find that half of the parameter space (with $0 < \alpha < 1$ and $0 <
\beta < 1$) was allowed until recently (at the expense of an
unnaturally high baryon fraction and scalar tilt).

{\bf Conclusions}. Using the recent measurements of temperature and
polarization anisotropies in the CMB by WMAP, as well as the matter
power spectrum measured by 2dFGRS, we obtained very stringent bounds
on a possible isocurvature component in the primordial
spectrum of density and velocity fluctuations. We have considered both
correlated and uncorrelated adiabatic and isocurvature modes, and
put strong constraints on the curvaton scenario of primordial 
perturbations.


\end{document}